\documentclass{IEEEtran}

\IEEEoverridecommandlockouts                              

\usepackage{graphics} 
\usepackage{epsfig} 
\usepackage{mathptmx} 
\usepackage{times} 
\usepackage{amsmath} 
\usepackage{amssymb}  
\usepackage{graphicx}

\usepackage{epstopdf}
\usepackage{epsfig}
\usepackage{enumerate}
\usepackage{caption2}
\newtheorem{Theorem 1}{Theorem}
\newtheorem{Theorem 2}[Theorem 1]{Theorem}
\newtheorem{Theorem 3}[Theorem 1]{Theorem}
\newtheorem{Theorem 4}[Theorem 1]{Theorem}
\newtheorem{Theorem 5}[Theorem 1]{Theorem}
\newtheorem{Definition 1}{Definition}
\newtheorem{Definition 2}[Definition 1]{Definition}
\newtheorem{Remark 1}{Remark}
\newtheorem{Remark 2}[Remark 1]{Remark}
\newtheorem{Lemma 1}{Lemma}

\DeclareSymbolFont{largesymbol}{OMX}{yhex}{m}{n}
\DeclareMathAccent{\Widehat}{\mathord}{largesymbol}{"62}

\title{\LARGE \bf Analysis and Design of Phase Desynchronization in Pulse-coupled Oscillators}

\author{Huan Gao and Yongqiang Wang
    \thanks{*The work was supported in part by the Institute for Collaborative Biotechnologies through grant W911NF-09-0001.}
    \thanks{Huan Gao and Yongqiang Wang are with the
        department of Electrical and Computer Engineering, Clemson
        University, Clemson, SC 29634, USA
        {\tt\small \{hgao2,yongqiw\}@clemson.edu}}%
}

\begin{document}

    \maketitle
    \thispagestyle{plain}
    \pagestyle{plain}
    \pagenumbering{arabic}

    \begin{abstract}

    By spreading phases on the unit circle,
    desynchronization algorithm is a powerful tool to achieve round-robin
    scheduling, which is crucial in applications as diverse as media
    access control of communication networks, realization of analog-to-digital
    converters, and scheduling of traffic flows in intersections.
    Driven by the increased application of pulse-coupled oscillators
    in achieving synchronization, desynchronization of pulse-coupled oscillators is also receiving more attention.
     In this paper, we propose a phase desynchronization algorithm by rigorously analyzing the dynamics of pulse-coupled oscillators and
     carefully designing the pulse based interaction function. A systematic proof for convergence to phase desynchronization  is
     also given. Different from many existing results which can only achieve equal separation of firing time instants, the proposed approach
     can achieve equal separation of phases, which is more difficult to achieve due to phase jumps in pulse-coupled oscillators. Furthermore, the new
     strategy can guarantee achievement of desynchronization even when some nodes
     have identical initial phases, a situation which  fails most existing desynchronization approaches. Numerical simulation results are
provided to illustrate the effectiveness of the theoretical results.

    \end{abstract}

    \section{Introduction}

     Pulse-coupled oscillators (PCOs) were originally proposed to model synchronization in biological
     systems
     such as flashing fireflies \cite{mirollo:90, goel2002synchrony} and firing neurons \cite{peskin:75,
     Ermentrout:96}. In recent years, with impressive scalability, simplicity, accuracy, and robustness, the PCO based synchronization strategy has become
 a powerful clock synchronization primitive for wireless sensor networks \cite{Simeone:08, Tyrrel:10, wernerallen:05, wang_TCST:12, TSP12}.

    A less explored property of pulse-coupled oscillators is desynchronization,
    which spreads the phase variables of all PCOs uniformly  apart (with equal difference between neighboring phases).
    Desynchronization has been found in many biological phenomena,
such as spiking neuron networks \cite{stopfer1997impaired} and the
communication signals of fish \cite{benda2006synchronization}.
What's more, desynchronization is also very important for Deep Brain
Stimulation (DBS) which has been proven an effective treatment for
Parkinson's disease \cite{nabi2010nonlinear}. Recently, phase
desynchronization has also been employed to perform time-division
multiple access (TDMA) medium access control (MAC) protocol
\cite{Degesys07,ashkiani2012pulse, taniguchi2013self}. Since
desynchronization enables all agents to send messages in a
round-robin manner, it provides collision-free message transmissions
and obtains a high throughput \cite{degesys2008self}.

    In the literature, a number of papers have emerged on PCO based desynchronization.
     Based on the PCO model in \cite{mirollo:90}, the authors in \cite{patel2007desynchronization} proposed a  desynchronization algorithm (INVERSE-MS) for an all-to-all network.
     The convergence property of  INVERSE-MS was further explored in
     \cite{pagliari2010bio} and \cite{phillips2013results,phillips2015robust}, using an algebraic framework and a hybrid systems framework, respectively.
     However, all the above results are about the achievement of  uniform firing
     time interval (equal time interval between any two consecutive PCOs' firings), which is referred to as weak desynchronization
\cite{patel2007desynchronization}, \cite{pagliari2010bio}. Weak
desynchronization relies on persistent phase jumps to achieve equal
time interval between consecutive firings, and hence  cannot
guarantee   uniformly spread of phases.

    Recently,   algorithms also emerged for phase desynchronization. The authors in \cite{degesys2007desync}
    proposed DESYNC in which each PCO relies on the firing information of the PCOs firing before and after this PCO when updating its phase variable.
     A modified algorithm V-DESYNC was proposed in \cite{settawatcharawanit2012v} to mitigate the beacon collision problem by randomly adding a small offset to the firing time.
     The authors in \cite{pagliari2010bio,rueetschi2011scheduling} proposed a  phase desynchronization algorithm by limiting the listening
     interval. The authors in   \cite{lien2012anchored} obtained phase desynchronization by adding an anchored PCO that never
adjusts its phase when other PCOs fire. Generally speaking,
performance of these desynchronization algorithms is very difficult
to analyze rigorously unless some kind of approximation is performed
first to simplify the PCO dynamics. Recently, a stochastic framework
is introduced to analyze the statistical behavior of the above
desynchronization algorithms
\cite{buranapanichkit2014stochastic,buranapanichkit2015convergence,deligiannis2015decentralized,deligiannis2015fast}.
However, such stochastic analysis assumes that PCO phase variables
 are subject to additive white noise, which is restrictive in
many applications. In fact, as pointed out in
\cite{lien2012anchored}, there is still a lack of rigorous
mathematical proof for the convergence of PCO based phase
desynchronization algorithms.

    In this paper, we propose a distributed PCO based phase desynchronization algorithm, and
    systematically characterize its convergence properties using rigorously  mathematical analysis.
    In addition, through designing the update rule we can achieve desynchronization even when some nodes have equal initial phase values,
    a situation which fails almost all existing PCO based desynchronization
    approaches.

    \section{PCO based phase desynchronization}

    In this section, we will first introduce the PCO model, and then we propose a phase desynchronization algorithm.

    \subsection{PCO model}

    We consider a network of $N$ PCOs with an all-to-all communication pattern. Each oscillator has a
    phase variable $\phi_k \in \mathbb{S}^1$ ($k=1,2,\ldots,N$) where $\mathbb{S}^1$ denotes the one-dimensional torus. Each phase variable $\phi_k$ evolves
    continuously from 0 to $2\pi$ with a constant speed determined by its natural
    frequency $w_k$. In this paper, the natural frequencies are assumed identical, i.e., $w_1=w_2=\ldots=w_N=w$.
    When an oscillator's phase reaches $2\pi$, it fires (emit a
    pulse) and resets its phase to $0$, after which the cycle
    repeats. When an oscillator receives a pulse from a neighboring
    oscillator, it shifts its phase by a certain amount   according to the phase response function, which is defined
    below:

    \begin{Definition 1}
        Phase response function $F(\phi_k)$ is defined as the phase shift (or jump) induced by a pulse as a function of  phase at which the pulse is received \cite{Izhikevich:07}.
    \end{Definition 1}

    Therefore, the interaction mechanism of PCOs can be described as follows:
    \begin{enumerate}
        \item Each PCO has a phase variable $\phi_k\in \mathbb{S}^1$ with initial value set to $\phi_k(0)$.
        $\phi_k$ evolves continuously from $0$ to $2\pi$ with a constant speed $w$;
        \item When the phase variable $\phi_k$ of PCO $k$ reaches $2\pi$, this
        PCO  fires, i.e., emits a pulse, and simultaneously resets  $\phi_k$ to 0.
        Then the same process repeats;
        \item  When a PCO receives a pulse from others, it updates its phase variable according to the phase response function
        $F(\phi_k)$:
        \begin{equation}\label{phase_update_sim1}
        \phi_k^{+}=\phi_k+F(\phi_k)
        \end{equation}
        where $\phi_k^{+}$ and $\phi_k$ denote the phases of the $k$th oscillator after and before a pulse.
    \end{enumerate}

    \subsection{PCO based phase desynchronization}

    It is already well-known that if the phase response function is chosen appropriately,
    pulse-coupled oscillators can achieve synchronization. For example, reference \cite{wang_tsp2:12}  shows that using a delay-advance
    phase response function in which the value of phase shift is negative in the interval $(0,\pi)$, positive in the interval $(\pi,2\pi)$, and zero at $0$ and $2\pi$,
    oscillator phases can achieve synchronization.

    Inspired by this idea, we propose a phase desynchronization algorithm by designing the phase response function of
    PCOs. Phase desynchronization in this paper is defined as follows:
    \begin{Definition 2}
        For a network of $N$ oscillators, phase desynchronization denotes the state on which all phase are distributed evenly on the unit
        circle with identical differences $\frac{2\pi}{N}$ between
        any two adjacent phases.
            \end{Definition 2}

    As discussed earlier, in PCO networks, phase desynchronization
    is more stringent than weak desynchronization \cite{patel2007desynchronization,pagliari2010bio} which uniformly
    spreads   firing time instants of constituent nodes. This is because in PCO
    networks, weak desynchronization can be realized by using
    persistent  phase jumps (caused by pulse interactions), which is
    not permitted by phase desynchronization; whereas weak synchronization follows naturally if phase desynchronization is achieved.

\begin{figure}[!h]
        \centering
        \includegraphics[width=0.35 \textwidth]{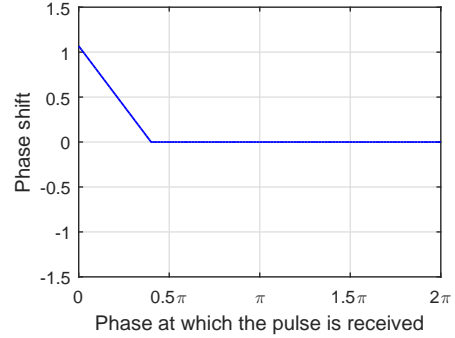}
        \caption{Phase response function ($\ref{eqn:PRC}$) for phase desynchronization algorithm based on PCOs ($N=5$, $l=0.85$).}
        \label{fig:4}
    \end{figure}

    The
    phase response function  $F(\phi_k)$ we propose is given by
    \begin{equation}\label{eqn:PRC}
    F(\phi_k)=\left\{\begin{aligned}
    & -l(\phi_k-\frac{2\pi} {N} ) &  {\quad} & 0\leq\phi_k< \frac{2\pi} {N} \\
    &\quad 0 & {\quad} & \frac{2\pi} {N}\leq\phi_k\leq2\pi
    \end{aligned}\right.
    \end{equation}
    where $0<l<1$. According to this phase response function, PCO $k$  updates
    its phase variable $\phi_k$ only when $\phi_k$ is within the interval $[0, \frac{2\pi} {N})$ as illustrated in Fig. \ref{fig:4}. Therefore, the phase update (\ref{phase_update_sim1}) can be rewritten as:
    \begin{equation}\label{phase_update}
    \phi_k^+=\left\{\begin{aligned}
    & (1-l)\phi_k+l \frac{2\pi} {N} &  {\quad} & 0\leq\phi_k< \frac{2\pi} {N} \\
    &\quad \phi_k & {\quad} & \frac{2\pi} {N}\leq\phi_k\leq2\pi
    \end{aligned}\right.
    \end{equation}
    \begin{Theorem 1}\label{Theorem_1}
        For a network of $N$ PCOs with no two PCOs having identical initial phases, the PCOs will achieve phase desynchronization
        if the phase response function $F(\phi_k)$ is given by (\ref{eqn:PRC}) for $0< l< 1$.
    \end{Theorem 1}

    Intuitively, under the phase response function (\ref{eqn:PRC}) which is positive in the interval $[0, \frac{2\pi} {N})$,
    an oscillator whose phase variable $\phi_k$ satisfies $\phi_k \in [0, \frac{2\pi} {N})$ will be pushed toward $\frac{2\pi} {N}$ when it receives a pulse from a firing oscillator
    (whose phase should be $2\pi$). Therefore,  the phase difference between oscillator $k$ and the oscillator who just fired will evolve towards $\frac{2\pi} {N}$.
     In the next section, we will provide a rigorous mathematical proof for Theorem $1$.

    \section{Convergence properties of the proposed phase desynchronization algorithm}

     In order to prove Theorem $\ref{Theorem_1}$, we need Theorem $\ref{Theorem_2}$ on the firing order of PCOs.
    \begin{Theorem 2}\label{Theorem_2}
        Under the phase response function (\ref{eqn:PRC}), the firing order of PCOs is time-invariant, i.e., if initially the $i$th oscillator
         fires immediately after the $j$th oscillator, then it will always
        fires immediately after the $j$th oscillator. In other words, the phase response function
        will not make one phase variable overpass another on $\mathbb{S}^1$.
    \end{Theorem 2}

    {\it Proof}: Assume that a phase variable $\phi_k$ reaches $2\pi$ and emits a pulse at
    time instant $t_k$. Then after receiving this pulse, all the other phase variables will
    update their values according to (\ref{eqn:PRC}). Suppose there are two phase
    variables satisfying $\phi_i<\phi_j$ before the pulse-induced update, and their values after update are
    denoted by $\phi_i^+$ and $\phi_j^+$, respectively. To prove theorem \ref{Theorem_2}, we only need
    to show that $\phi_i^+ < \phi_j^+$ is always true. Because the values of $\phi_i$ and $\phi_j$
    are changed by pulses only when they are less than
    $\frac{2\pi}{N}$, depending on the relationship between
    $\phi_i$, $\phi_j$, and $\frac{2\pi}{N}$, we divide the analysis
    into the following three cases:
    \begin{enumerate}
        \item If $\phi_i \geq \frac{2\pi} {N}$ and  $\phi_j \geq \frac{2\pi} {N}$ hold, then the values of
        both $\phi_i$ and $\phi_j$ will not be
        affected by the pulse according to (\ref{eqn:PRC}). So it follows naturally
        that $\phi_i^+<\phi_j^+$
        is still true after the pulse;
        \item If $\phi_i < \frac{2\pi} {N}$ and  $\phi_j < \frac{2\pi} {N}$ hold,
        then the values of both $\phi_i$ and $\phi_j$ will be changed by the pulse.
        According to the phase update (\ref{phase_update}), they will become
        \begin{equation}
        \phi_i ^+=(1-l)\phi_i+l \frac{2\pi} {N}
        \end{equation}
        and
        \begin{equation}
        \phi_j ^+=(1-l)\phi_j+l \frac{2\pi} {N}
        \end{equation}
        respectively. It can be verified that $\phi_i^+<\phi_j^+$ is still true if before the pulse the condition  $\phi_i<\phi_j$ holds;
        \item  If $\phi_i < \frac{2\pi} {N}$ and  $\phi_j \geq \frac{2\pi} {N}$ are true,
        then only the value of $\phi_i$ will be affected by the pulse.
        So we have $\phi_j^+=\phi_j \geq \frac{2\pi}{N}$ and $\phi_i ^+=(1-l)\phi_i+l \frac{2\pi} {N}$,
        which is still less than $\frac{2\pi}{N}$ since
        \begin{equation}
        \begin{aligned}
        \phi_i ^+-\frac{2\pi}{N}&=(1-l)\phi_i+l \frac{2\pi}{N}-\frac{2\pi}{N}\\
        &=(\phi_i-\frac{2\pi}{N})(1-l)\\
        & < 0
        \end{aligned}
        \end{equation}
        Therefore, we have $\phi_i ^+ < \frac{2\pi}{N}$ and
        $\phi_j^+ \geq \frac{2\pi}{N}$, which means that $\phi_i ^+<\phi_j
        ^+$ is still true.
    \end{enumerate}

    In conclusion, $\phi_i^+ < \phi_j^+$ will always be true if $\phi_i < \phi_j$ holds, which means that the phase response function (\ref{eqn:PRC})
    will not change the firing order of all the PCOs. \hfill$\blacksquare$

\begin{figure}[!h]
        \centering
        \includegraphics[width=0.35 \textwidth]{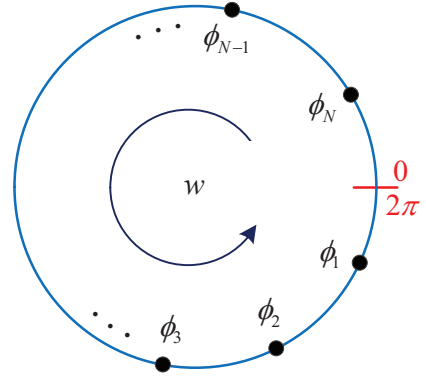}
        \caption{Initially (at $t=0$), the phases of PCOs are arranged in a way such that $\phi_1(0) > \phi_2(0) > \ldots > \phi_N(0)$ holds.}
        \label{fig:02}
    \end{figure}

    To prove Theorem \ref{Theorem_1} we also need an index to measure the degree of achievement of desynchronization. Without loss of generality, we denote the initial time instant as $t=0$ and assume at $t=0$ that the phases of PCOs are
    arranged in a way such that $\phi_1(0) > \phi_2(0) > \ldots > \phi_N(0)$
    holds, as illustrated in Fig. \ref{fig:02}. (Note that here we assume that no two PCOs' initial phases are identical. The assumption will be relaxed in Sec. IV.)
    From Theorem $\ref{Theorem_2}$, we know that the firing order of PCOs will not be affected by the pulse-induced update. So if $\phi_k$ is the immediate follower (anti-clockwisely) of $\phi_{k-1}$  on $\mathbb{S}^1$ at $t=0$, it
    will always be the immediate follower (anti-clockwisely) of
    $\phi_{k-1}$ on $\mathbb{S}^1$.
    Therefore, the phase differences between neighboring
    PCOs can always be expressed as:
    \begin{equation}\label{eq:phase difference}
    \left\{ \begin{aligned}
    \Delta_k&=(\phi_k-\phi_{k+1} ) \ {\rm mod}  \ 2\pi , \quad k=1,2,\ldots,N-1\\
    \Delta_N&=(\phi_N-\phi_1) \ {\rm mod} \ 2\pi
    \end{aligned}\right.
    \end{equation}

    As we discussed earlier, phase desynchronization is defined as that the phase variables of all PCOs are uniformly spread apart.
    In other words, all the phase differences between neighboring (in terms of phase) PCOs are equal to $\frac{2\pi} {N}$.
    Therefore, for the convenience in analysis, we introduce an index $P$ to measure the
    degree of achievement of phase desynchronization by using the phase differences between
     neighboring (in terms of phase) PCOs:
    \begin{equation}\label{eq:index_desync}
    P\triangleq \sum\limits_{k=1}^{N} |\Delta_k-\frac{2\pi}{N}|
    \end{equation}
    When phase desynchronization is achieved, the phase differences between neighboring
    PCOs are equal to $\frac{2\pi}{N}$, so the index $P$ in (\ref{eq:index_desync}) will
     reach its minimum $0$.
    It can also be easily verified that $P$ equals $0$ only when phase desynchronization is achieved.

    Therefore, in order to prove Theorem $\ref{Theorem_1}$, we only need to prove that the index $P$ will converge to $0$ under the phase response function (\ref{eqn:PRC}).

    {\it Proof of Theorem \ref{Theorem_1}}: From the analysis above, to prove Theorem
    $\ref{Theorem_1}$,
    we only need to prove that the index $P$ will converge to $0$. According to the  interaction mechanism of PCOs in Sec. II-A, all
phase variables will evolve towards $2\pi$ with the same speed $w$,
thus the phase differences between neighboring PCOs will not change
during  two consecutive pulses, neither will $P$. Therefore we only
need to concentrate on how $P$ evolves at discrete-time instants
when pulses are emitted.

   After $t=0$, according to the dynamics of PCOs, all phase variables
    evolve towards $2\pi$ with the same speed
   $w$. Since $\phi_1$ is the largest, it will  reach $2\pi$ first  without perturbation (denote this  time instant as $t=t_1$), upon which PCO $1$ will send a pulse which will be received by all the other PCOs. After receiving the pulse, PCO $i$  will update $\phi_i$ ($2\leq i\leq N$) according to (\ref{eqn:PRC}). From
   (\ref{eqn:PRC}), we know that the value of $\phi_i$ will be changed if it is within the interval $[0,\frac{2\pi}{N})$;
    otherwise, it will keep unchanged. Therefore we call the interval $[0,\frac{2\pi}{N})$ as the ``effective
    interval." If at least one phase variable is within the ``effective interval,''
    then this phase variable  will be affected by the pulse, which will in turn affect the
    phase differences between neighboring (in terms of phase) PCOs. In this case the pulse  will be called an ``active pulse.''  Otherwise, if no $\phi_i$  ($2\leq i\leq N$) are within the ``effective interval,'' then the pulse from PCO $1$ will not
   affect any phase variable and it will be referred to as a ``silent pulse.''
   If the pulse from PCO $1$ is a ``silent pulse,'' other than resetting $\phi_1$ from $2\pi$
   to $0$, it will not disturb the evolution of any other phase variable
   towards $2\pi$. After $t=t_1$, $\phi_1$ becomes the smallest and $\phi_{2}$ becomes the largest who will reach $2\pi$ next.
   Suppose $\phi_{2}$ reaches $2\pi$ at time instant $t=t_2$ and emits a pulse.
   Similar to the pulse from PCO $1$, this pulse could be an ``active pulse'' or a ``silent
   pulse.'' Following the same line of reasoning, it can be inferred that $\phi_3$
   will reach $2\pi$ after $\phi_2$'s resetting and its pulse can be an ``active
   pulse'' or a ``silent pulse.'' So do the other $\phi_i$s and their
   corresponding pulses. Next we prove that there cannot be $N$ consecutive
    ``silent pulses'' unless phase desynchronization is achieved.

    We use proof of contradiction. Assume that $N$ consecutive pulses are all
    ``silent pulses'' but phase desynchronization has not been achieved. From Theorem $\ref{Theorem_2}$, no phase variable can surpass another on $\mathbb{S}^1$,
    so the $N$ consecutive pulses
    must be from $N$ different PCOs. For a pulse from oscillator $i$ to be a ``silent pulse,'' the phase variable of the PCO who sents a pulse immediately before oscillator
     $i$ (whose phase is the smallest according to Theorem $\ref{Theorem_2}$) must be no less than $\frac{2\pi}{N}$ (outside the ``effective interval'').
     Therefore, if there are $N$ consecutive ``silent pulses,'' then the phase difference between any two neighboring PCOs is no less than $\frac{2\pi}{N}$.
      Because the sum of all the phase differences are $2\pi$, we can infer that all the phase differences are equal to $\frac{2\pi}{N}$, meaning that phase desynchronization is achieved, which contradicts the initial assumption.

\begin{figure}[!h]
        \centering
        \includegraphics[width=0.45 \textwidth]{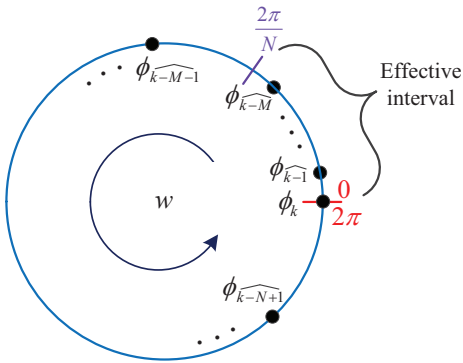}
        \caption{The phase variables $\phi_{\Widehat{k-1}}, \ldots, \phi_{\Widehat{k-M}}$ are in the  ``effective interval'' when oscillator $k$ sends the first ``active pulse'' at $t=t_k$.}
        \label{fig:03}
    \end{figure}

     Therefore, there cannot be $N$ consecutive ``silent pulses'' unless phase desynchronization is achieved.
   Without loss of generality, we assume that the pulse from oscillator $k$ is the first
   ``active pulse,'' and its phase variable $\phi_k$ reaches $2\pi$ at time instant $t=t_k$. Since the pulse is an ``active
    pulse,'' there is at least one phase variable within the ``effective interval'' when the pulse is
     sent. Without loss of generality, we assume that there are $M$
     ($M$ is an integer satisfying $1\leq M\leq N-1$)
     phase variables within the ``effective interval'' which can be represented as $\phi_{\Widehat{k-1}}, \ldots, \phi_{\Widehat{k-M}}$,
     where the superscript `` $\Widehat{}$ '' represents modulo operation on $N$, i.e.,
     $\Widehat{\bullet}\triangleq(\bullet) \ {\rm mod} \ N$, as illustrated in Fig.
     \ref{fig:03}. According to the assumption, we have  $\phi_{\Widehat{k-M}}<
     \frac{2\pi}{N}\leq\phi_{\Widehat{k-M-1}}$.
     Since $\phi_{\Widehat{k-1}}, \ldots, \phi_{\Widehat{k-M}}$ are in the ``effective interval,'' they will update their values after receiving the pulse from oscillator
      $k$ according to the phase update in (\ref{phase_update}) as follows:
    \begin{equation}\label{eq:phase update}
    \phi_{\Widehat{k-i}} ^+=(1-l)\phi_{\Widehat{k-i}} +l \frac{2\pi}
    {N},\quad i=1,\ldots,M
    \end{equation}
     Note that we also have   $\phi_k ^+=0$ and $\phi_{\Widehat{k-j}}  ^+=\phi_{\Widehat{k-j}} $ for $j=M+1,\ldots,N-1$
     (because
     $\phi_{\Widehat{k-j}}$ for $j=M+1,\ldots,N-1$  are not in the ``effective
     interval'' and thus will not be changed according to the phase response function in (\ref{eqn:PRC})).

    Therefore, phase differences after the update can be rewritten as follows:
    \begin{equation}\label{eq:phase difference update}
    \left\{ \begin{aligned}
    &\Delta_{k} ^+=\phi_{k }^+ -\phi_{\Widehat{k-N+1}} ^+ +2\pi= 2\pi-\phi_{\Widehat{k-N+1}}\\
    &\Delta_{\Widehat{k-1}} ^+=\phi_{\Widehat{k-1}} ^+-\phi_{k} ^+= (1-l)\phi_{\Widehat{k-1}} +l \frac{2\pi} {N}  \\
    &\Delta_{\Widehat{k-i}} ^+=\phi_{\Widehat{k-i}} ^+-\phi_{\Widehat{k-i+1}} ^+=(1-l)(\phi_{\Widehat{k-i}}-\phi_{\Widehat{k-i+1}}), i=2,\ldots,M \\
    &\Delta_{\Widehat{k-M-1}} ^+=\phi_{\Widehat{k-M-1}} ^+-\phi_{\Widehat{k-M}} ^+=\phi_{\Widehat{k-M-1}}- (1-l)\phi_{\Widehat{k-M}} -l \frac{2\pi} {N}  \\
    &\Delta_{\Widehat{k-j}} ^+=\phi_{\Widehat{k-j}} ^+-\phi_{\Widehat{k-j+1}} ^+= \phi_{\Widehat{k-j}} -\phi_{\Widehat{k-j+1}}, j=M+2,\ldots,N-1\\
    \end{aligned}\right.
    \end{equation}

    According to (\ref{eq:index_desync}), the new $P$ (denote it as $P ^+$) after the update is given by:
    \begin{equation}\label{eq:P update}
    P ^+=\sum\limits_{j=0}^{N-1} |\Delta_{\Widehat{k-j}} ^+  -  \frac{2\pi} {N}|
    \end{equation}

    To show that the pulse from oscillator $k$ will decrease  $P$, we calculate the difference
    of $P$ before and after the pulse-induced update:
    \begin{equation}\label{eq:P change}
    \begin{aligned}
    P ^+-P &= \sum\limits_{j=0}^{N-1} |\Delta_{\Widehat{k-j}} ^+  -  \frac{2\pi} {N}|-\sum\limits_{j=0}^{N-1} |\Delta_{\Widehat{k-j}}   -  \frac{2\pi}{N}|\\
    &=\underbrace{|\Delta_{k} ^+  -  \frac{2\pi} {N}|-|\Delta_{k}-\frac{2\pi} {N}|}_{\rm part\, 1}+\underbrace{|\Delta_{\Widehat{k-1}} ^+  -  \frac{2\pi} {N}|-|\Delta_{\Widehat{k-1}}-\frac{2\pi} {N}|}_{\rm
    part\, 2}\\
    & \quad +\underbrace{\sum\limits_{i=2}^{M} |\Delta_{\Widehat{k-i}} ^+-\frac{2\pi} {N}|-\sum\limits_{i=2}^{M} |\Delta_{\Widehat{k-i}} - \frac{2\pi} {N}|}_{\rm part
    \,3}\\
    & \quad +\underbrace{|\Delta_{\Widehat{k-M-1}} ^+-\frac{2\pi} {N}|-|\Delta_{\Widehat{k-M-1}} -\frac{2\pi} {N}|}_{\rm part\,4}\\
    & \quad +\underbrace{\sum\limits_{j=M+2}^{N-1} |\Delta_{\Widehat{k-j}} ^+-\frac{2\pi} {N}|-\sum\limits_{j=M+2}^{N-1} |\Delta_{\Widehat{k-j}} - \frac{2\pi} {N}|}_{\rm part
    \,5}
    \end{aligned}
    \end{equation}
    The  part 1, part 2, part 3, and part 5  in  (\ref{eq:P change}) can be simplified as follows, respectively:

    \begin{enumerate}
        \item
        \begin{equation}
        \begin{aligned}
        &\underbrace{|\Delta_{k} ^+  -  \frac{2\pi} {N}|-|\Delta_{k}-\frac{2\pi} {N}|}_{\rm part\,1}\\
        =&|2\pi-\phi_{\Widehat{k-N+1}} -\frac{2\pi} {N}|-|2\pi-\phi_{\Widehat{k-N+1}} -\frac{2\pi} {N}|\\
        =&\ 0
        \end{aligned}
        \end{equation}
        \item
        \begin{equation}
        \begin{aligned}
        &\underbrace{|\Delta_{\Widehat{k-1}} ^+  -  \frac{2\pi} {N}|-|\Delta_{\Widehat{k-1}}-\frac{2\pi} {N}|}_{\rm part\,2}\\
        =&| (1-l)\phi_{\Widehat{k-1}} +l \frac{2\pi} {N} -\frac{2\pi} {N}|-|\phi_{\Widehat{k-1}} -\frac{2\pi} {N}|\\
        =&(1-l)(\frac{2\pi} {N}-\phi_{\Widehat{k-1}}) -(\frac{2\pi} {N}-\phi_{\Widehat{k-1}})\\
        =&-l(\frac{2\pi} {N}-\phi_{\Widehat{k-1}})
        \end{aligned}
        \end{equation}
        In the above derivation we used $\phi_{\Widehat{k-1}}<\frac{2\pi} {N}$.
        \item
        \begin{equation}
        \begin{aligned}
        &\underbrace{\sum\limits_{i=2}^{M} |\Delta_{\Widehat{k-i}} ^+-\frac{2\pi} {N}|-\sum\limits_{i=2}^{M} |\Delta_{\Widehat{k-i}} - \frac{2\pi} {N}|}_{\rm part\,3}\\
        =&\sum\limits_{i=2}^{M} \{|(1-l)(\phi_{\Widehat{k-i}}-\phi_{\Widehat{k-i+1}})  -  \frac{2\pi} {N}|\\
        & \quad - |(\phi_{\Widehat{k-i}}-\phi_{\Widehat{k-i+1}})   -  \frac{2\pi} {N}|\}\\
        =&\sum\limits_{i=2}^{M} \{\frac{2\pi} {N}-(1-l)(\phi_{\Widehat{k-i}}-\phi_{\Widehat{k-i+1}})-\frac{2\pi} {N}\\
        &  \quad + (\phi_{\Widehat{k-i}}-\phi_{\Widehat{k-i+1}})\}\\
        =&\sum\limits_{i=2}^{M} l(\phi_{\Widehat{k-i}}-\phi_{\Widehat{k-i+1}})
        \end{aligned}
        \end{equation}
        where we used the relationships $\phi_{\Widehat{k-i}}-\phi_{\Widehat{k-i+1}}<\frac{2\pi} {N}$ and $(1-l)(\phi_{\Widehat{k-i}}-\phi_{\Widehat{k-i+1}})<\frac{2\pi} {N}$, $i=2,\ldots,M$.
        \item
        \begin{equation}\label{eq:item 5}
        \begin{aligned}
        &\underbrace{\sum\limits_{j=M+2}^{N-1} |\Delta_{\Widehat{k-j}} ^+-\frac{2\pi} {N}|-\sum\limits_{j=M+2}^{N-1} |\Delta_{\Widehat{k-j}} - \frac{2\pi} {N}|}_{\rm part\,5}\\
        =&\sum\limits_{j=M+2}^{N-1} \{|\phi_{\Widehat{k-j}} -\phi_{\Widehat{k-j+1}}-\frac{2\pi} {N}|- |\phi_{\Widehat{k-j}} -\phi_{\Widehat{k-j+1}} -\frac{2\pi} {N}| \} \\
        =&\ 0
        \end{aligned}
        \end{equation}
     \end{enumerate}

    Combining  (\ref{eq:P change})-(\ref{eq:item 5}) leads to:
    \begin{equation}\label{eq:P change simplification}
    \begin{aligned}
    P ^+-P=&-l(\frac{2\pi} {N}-\phi_{\Widehat{k-1}})+\sum\limits_{i=2}^{M} l(\phi_{\Widehat{k-i}}-\phi_{\Widehat{k-i+1}})\\
    & \quad +|\Delta_{\Widehat{k-M-1}} ^+-\frac{2\pi} {N}|-|\Delta_{\Widehat{k-M-1}} -\frac{2\pi} {N}|\\
    =& l\phi_{\Widehat{k-M}} -l \frac{2\pi} {N} +|\Delta_{\Widehat{k-M-1}} ^+-\frac{2\pi} {N}|-|\Delta_{\Widehat{k-M-1}} -\frac{2\pi} {N}|
    \end{aligned}
    \end{equation}

    Next, we discuss the value of $P^+-P$ in (\ref{eq:P change simplification}) under three different cases:
    \begin{enumerate}[{Case} 1{:}]
        \item
        If $\Delta_{\Widehat{k-M-1}}>\frac{2\pi} {N}$ and
        $\Delta_{\Widehat{k-M-1}} ^+ \geq \frac{2\pi} {N}$ hold, (\ref{eq:P change simplification}) can be rewritten as:
        \begin{equation}\label{eq:P case 1}
        \begin{aligned}
        P ^+-P=& l\phi_{\Widehat{k-M}} -l \frac{2\pi} {N} +\Delta_{\Widehat{k-M-1}} ^+-\Delta_{\Widehat{k-M-1}} \\
        =& l\phi_{\Widehat{k-M}} -l \frac{2\pi} {N}+\phi_{\Widehat{k-M-1}}- (1-l)\phi_{\Widehat{k-M}} \\
        & \quad -l \frac{2\pi} {N} -\phi_{\Widehat{k-M-1}}+\phi_{\Widehat{k-M}}\\
        = & 2l (\phi_{\Widehat{k-M}}- \frac{2\pi} {N})\\
        < & 0
        \end{aligned}
        \end{equation}
        \item    If $\Delta_{\Widehat{k-M-1}} >\frac{2\pi} {N}$ and
        $\Delta_{\Widehat{k-M-1}} ^+< \frac{2\pi} {N}$ hold, we have
        $\phi_{\Widehat{k-M}}-\phi_{\Widehat{k-M-1}}+\frac{2\pi} {N}<0$.
        Then (\ref{eq:P change simplification}) can be rewritten as:
        \begin{equation}\label{eq:P case 2}
        \begin{aligned}
        P ^+-P  = & l\phi_{\Widehat{k-M}} -l \frac{2\pi} {N}+ \frac{2\pi} {N} -\Delta_{\Widehat{k-M-1}} ^+  -\Delta_{\Widehat{k-M-1}}\\
        & \quad + \frac{2\pi} {N}\\
        =& l\phi_{\Widehat{k-M}} -l \frac{2\pi} {N}+ \frac{2\pi} {N} -\phi_{\Widehat{k-M-1}}+ (1-l)\phi_{\Widehat{k-M}}\\
        & \quad  +l \frac{2\pi} {N}-\phi_{\Widehat{k-M-1}}+\phi_{\Widehat{k-M}} + \frac{2\pi} {N}\\
        =& 2(\phi_{\Widehat{k-M}}-\phi_{\Widehat{k-M-1}}+\frac{2\pi} {N}) \\
        <& 0
        \end{aligned}
        \end{equation}
        \item     If $\Delta_{\Widehat{k-M-1}}\leq\frac{2\pi} {N}$ and
        $\Delta_{\Widehat{k-M-1}} ^+ < \frac{2\pi} {N}$ hold, (\ref{eq:P change simplification}) can be rewritten as:
        \begin{equation}\label{eq:P case 3}
        \begin{aligned}
        P ^+-P  =& l\phi_{\Widehat{k-M}} -l \frac{2\pi} {N} -\Delta_{\Widehat{k-M-1}} ^+ +\Delta_{\Widehat{k-M-1}} \\
        =& l\phi_{\Widehat{k-M}} -l \frac{2\pi} {N}-\phi_{\Widehat{k-M-1}}+ (1-l)\phi_{\Widehat{k-M}} \\
        &  \quad +l \frac{2\pi} {N} +\phi_{\Widehat{k-M-1}}-\phi_{\Widehat{k-M}}\\
        =& 0
        \end{aligned}
        \end{equation}
    \end{enumerate}

    \begin{Remark 1}
        According to (\ref{eqn:PRC}), we cannot have a
        fourth case where $\Delta_{\Widehat{k-M-1}}\leq\frac{2\pi} {N}$
        and $\Delta_{\Widehat{k-M-1}} ^+ \geq \frac{2\pi} {N}$ hold
        because of the following constraint according to (\ref{eq:phase difference}) and (\ref{eq:phase difference update}):
        \begin{equation}\label{eq:remark 2}
        \begin{aligned}
        & \Delta_{\Widehat{k-M-1}} ^+-\Delta_{\Widehat{k-M-1}}\\
        =& \phi_{\Widehat{k-M-1}}- (1-l)\phi_{\Widehat{k-M}} -l \frac{2\pi} {N}-\phi_{\Widehat{k-M-1}}+\phi_{\Widehat{k-M}}\\
        =& l(\phi_{\Widehat{k-M}}-\frac{2\pi} {N})\\
        <& 0
        \end{aligned}
        \end{equation}
        It is worth noting that in (\ref{eq:remark 2}) we used the initial assumption $\phi_{\Widehat{k-M}} < \frac{2\pi} {N}$.
    \end{Remark 1}

    From the above analysis, we know that the value of $P$ will be decreased or  unchanged by each ``active pulse.'' Next we
    proceed to show that $P$ will not be retained at a non-zero value, i.e., the ``Case 3"
    above cannot always be true before phase desynchronization is achieved.

\begin{figure}
        \centering
        \includegraphics[width=0.5\textwidth]{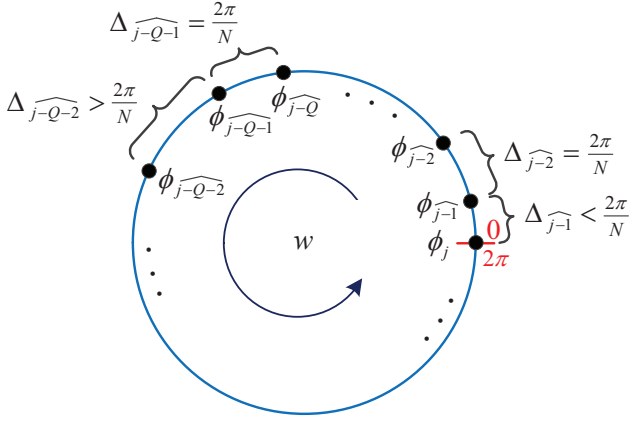}
        \caption{The ``state one'' that $Q$ ($0 \leq Q \leq N-2$) phase differences between the larger phase difference and the smaller phase difference are equal to $\frac{2\pi} {N}$ .}
        \label{fig:021}
    \end{figure}

\begin{figure}
        \centering
        \includegraphics[width=0.5\textwidth]{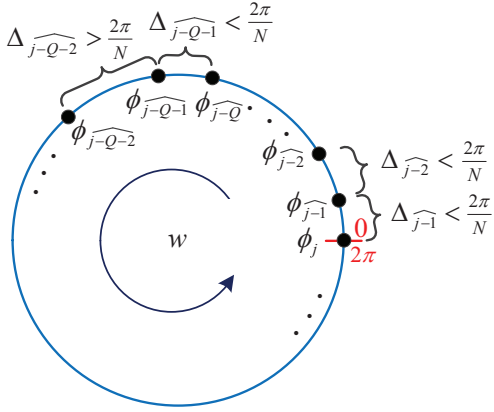}
        \caption{The ``state two'' that $Q$ ($0 \leq Q \leq N-2$) phase differences between the larger phase difference and the smaller phase difference become smaller than $\frac{2\pi} {N}$. }
        \label{fig:022}
    \end{figure}

    It can be easily inferred that before achieving the phase desynchronization
    there always exists one phase difference larger than $\frac{2\pi} {N}$ and one phase
     difference  smaller than $\frac{2\pi} {N}$, and in between the two phase differences
     there may be some phase differences (represent the number as $Q$, $0 \leq Q \leq N-2$) that
      are equal to $\frac{2\pi} {N}$, which is defined as  ``state one.'' Denote the phase
      difference larger than $\frac{2\pi} {N}$ and the phase difference smaller than
      $\frac{2\pi} {N}$   as $\Delta_{\Widehat{j-Q-2}}$ and $\Delta_{\Widehat{j-1}}$, respectively,
       and    the  $Q$  phase differences (which are equal to $\frac{2\pi} {N}$) in between as
        $\Delta_{\Widehat{j-2}}, \ldots, \Delta_{\Widehat{j-Q-1}}$ (cf. Fig. \ref{fig:021}).
         It can be proven that ``state one'' must evolve to a ``state two''
         (cf. Fig. \ref{fig:022}) after $Q(N-1)$  pulses. In the
         ``state
         two,'' the
          $Q$ phase differences which were equal to $\frac{2\pi} {N}$ in the ``state one''
          become smaller than $\frac{2\pi} {N}$, meaning that the condition in  ``Case 3"
          is not satisfied when oscillator $\Widehat {j-Q}$ fires because phase $\phi_{\Widehat{j-Q-1}}$ is within the ``effective interval'' and $\Delta_{\Widehat{k-M-1}}=\Delta_{\Widehat{j-Q-2}}>\frac{2\pi} {N}$ ($k=\Widehat{j-Q}$ and $M=1$) is true.


\begin{figure}
        \centering
        \includegraphics[width=0.50\textwidth]{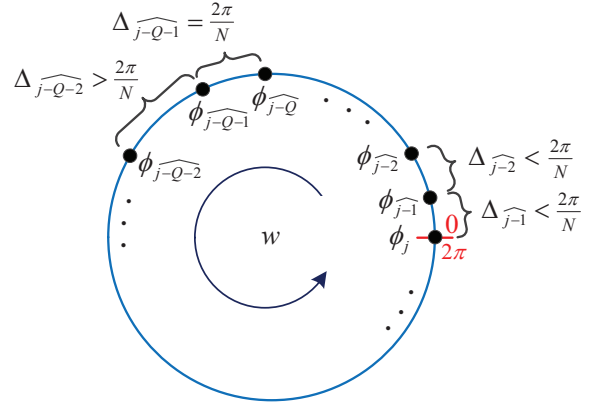}
        \caption{ $Q-1$ ($0 \leq Q \leq N-2$) phase differences between the larger phase difference and the smaller phase difference  are equal to $\frac{2\pi} {N}$. }
        \label{fig:023}
    \end{figure}

    Now we will illustrate how those phase differences equal to $\frac{2\pi} {N}$ become
    smaller than  $\frac{2\pi} {N}$ after $Q(N-1)$ firing. Suppose at $t=t_j$,
    an ``active pulse'' is emitted by oscillator $j$. This pulse only affects $\phi_{\Widehat{j-1}}$
    since only $\phi_{\Widehat{j-1}}$ is within the ``effective interval,'' and it
    increases the value of  $\phi_{\Widehat{j-1}}$
    ($\phi_{\Widehat{j-1}} ^+ - \phi_{\Widehat{j-1}} =l(\frac{2\pi} {N}-\phi_{\Widehat{j-1}})>0$),
     which in turn makes $\Delta_{\Widehat{j-2}}$ smaller than  $\frac{2\pi} {N}$.
      As time evolves, oscillators $\Widehat{j+1},\ldots,\Widehat{j-Q-2}$ will fire one by one.
      To discuss the evolution of PCOs under these pulses, we need to
      distinguish between  two different cases. The first case is that
       $\phi_{\Widehat{j-1}}$ is still within the ``effective interval''
       which means that the pulses will make $\Delta_{\Widehat{j-2}}$  smaller and smaller.
        The second case is that $\phi_{\Widehat{j-1}}$ is not within the ``effective interval''
         and thus it is not affected by those pulses. Therefore,  $\Delta_{\Widehat{j-2}}$
         keeps unchanged and is still smaller than  $\frac{2\pi} {N}$ as we discussed above.
          Both cases render the same result that $\Delta_{\Widehat{j-2}}$ is smaller than
            $\frac{2\pi} {N}$ after those pulses.
            Next oscillators $\Widehat{j-Q-1},\ldots,\Widehat{j-2}$ will fire one after another.
            However, their pulses are all ``silent pulses'' since no phase variables
            are within the ``effective interval.'' So
            all the phase differences will keep unchanged, meaning that  $\Delta_{\Widehat{j-Q-2}}$ is
            still larger than $\frac{2\pi} {N}$, $\Delta_{\Widehat{j-Q-1}}, \ldots, \Delta_{\Widehat{j-3}}$
            are equal to $\frac{2\pi} {N}$, and $\Delta_{\Widehat{j-2}}$ is smaller than
            $\frac{2\pi} {N}$. Therefore, after $N-1$ consecutive  firing,
            the number of phase differences equal to $\frac{2\pi} {N}$ is reduced by one to $Q-1$,
            as illustrated in Fig. (\ref{fig:023}).
            Therefore, after $Q(N-1)$ firing, the phase differences
             equal to $\frac{2\pi} {N}$ in the ``state one'' become smaller than  $\frac{2\pi} {N}$,
             which means that the ``state two'' is achieved.

    So the ``state one'' must evolve to the ``state two,''
    and thus the condition in  ``Case 3" cannot always exist before achieving
    phase desynchronization because $\Delta_{\Widehat{k-M-1}}>\frac{2\pi} {N}$ ($k=\Widehat{j-Q}$ and $M=1$)
    will be true when PCO $\Widehat {j-Q}$ fires after the ``state two'' is achieved.
    Consequently, $P$  will keep decreasing until it  reaches $0$, i.e., until phase
     desynchronization is achieved. Therefore, the PCOs will achieve
      phase desynchronization under the phase response function (\ref{eqn:PRC}) for $0< l< 1$. \hfill$\blacksquare$

    \section{Dealing with oscillators with identical phases}

    In Sec. II and Sec. III, we achieved phase desynchronization for a network of $N$ PCOs
    if no two oscillators have identical initial phases. In fact, under the proposed
    mechanism, if there are some oscillators
    (represent the number as $X$, $2 \leq X \leq N$) having equal initial phases,
    these $X$ PCOs will always have equal phases. This is because these $X$ PCOs will
    always make updates simultaneously  with identical phase shifts.
  Therefore, these $X$ PCOs will become inseparable, making phase desynchronization
  impossible.

    We propose the following modifications to the original phase
    update rule to address this issue: When an oscillator's phase reaches
    $2\pi$, this oscillator resets its phase to different values depending on whether
    a phase  is detected. More specifically, if at this time instant, the oscillator
    also detect a pulse from its neighbor (meaning that this oscillator has equal phase with
    the neighbor), it will reset its phase to a value randomly chosen from $[0, 2\pi)$;
    otherwise it will still reset its phase to 0.

    Therefore, the interaction mechanism of PCOs will
    become:
    \begin{enumerate}
        \item Each PCO has a phase variable $\phi_k\in \mathbb{S}^1$ with initial value
        set to $\phi_k(0)$.
        $\phi_k$ evolves continuously from $0$ to $2\pi$ with a constant speed $w$;
        \item When the phase variable $\phi_k$ of PCO $k$ reaches $2\pi$, this
        PCO  fires, i.e., emits a pulse, and simultaneously resets  $\phi_k$ to $\phi_{k,0}$ whose value
        depends
        on whether a pulse from a neighbor is detected: 1) if no
        pulses from neighbors are detected, then $\phi_{k,0}=0$; and 2)
        if a pulse from neighbors is detected, then $\phi_{k,0}$
        will be a value randomly chosen from the interval
        $[0,\,2\pi)$.
        Then the same process repeats;
        \item When a PCO receives a pulse from others, it updates its phase variable according to the phase response function
        $F(\phi_k)$:
        \begin{equation}\label{phase_update_sim}
        \phi_k^{+}=\phi_k+F(\phi_k)
        \end{equation}
        where $\phi_k^{+}$ and $\phi_k$ denote the phases of the $k$th oscillator after and before the pulse.

    \end{enumerate}

    Under this new mechanism, oscillators with identical phases will
    be separated and hence desynchronization can also be achieved even some oscillator have equal initial phase values, which will be confirmed by numerical simulations in Sec. V.

    \section{Simulation results}

\begin{figure}
        \centering
        \includegraphics[width=0.5\textwidth]{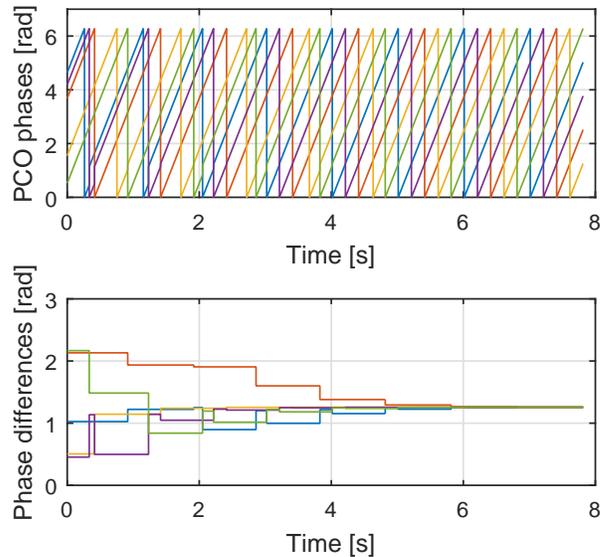}
        \caption{The evolution of PCO phases $\phi_k (k=1,\ldots,N)$ (upper panel) and phase differences between neighboring PCOs $\Delta_k (k=1,\ldots,N)$ (lower panel)
        under the phase response function $(\ref{eqn:PRC})$. The initial phase values were randomly chosen from the interval  $[0,\,2\pi)$. }
        \label{fig:7}
    \end{figure}

\begin{figure}
        \centering
        \includegraphics[width=0.42\textwidth]{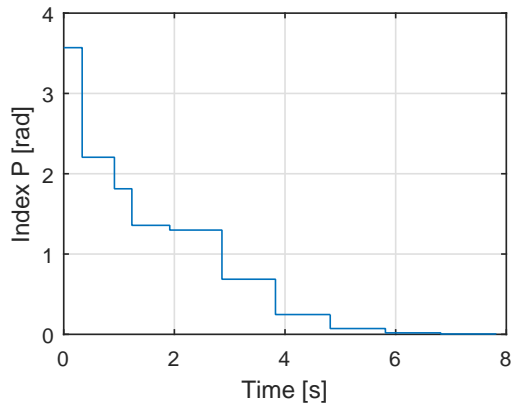}
        \caption{The evolution of   index $P$ under the phase response function $(\ref{eqn:PRC})$. The initial phase values were randomly chosen from the interval  $[0,\,2\pi)$. }
        \label{fig:9}
    \end{figure}

    We use simulation results to demonstrate the proposed phase desynchronization
    algorithm.
    We first considered the case where no two oscillators have equal initial phases.
     The initial phases were randomly chosen from $[0,2\pi)$ and  $l$ in the phase response function $(\ref{eqn:PRC})$
     was set to $0.85$.  $w$ was set to $2\pi$. The evolutions of oscillator phases,
      phase differences between neighboring PCOs, and the index $P$ are given in Fig. $\ref{fig:7}$
       and Fig. $\ref{fig:9}$, respectively. It can be seen that
       the PCO phases were uniformly spread apart,   the phase differences between
       neighboring PCOs converged to $\frac{2\pi} {5}$, and the index $P$ converges to $0$.
       This confirmed the effectiveness of the proposed desynchronization algorithm.

\begin{figure}
        \centering
        \includegraphics[width=0.5\textwidth]{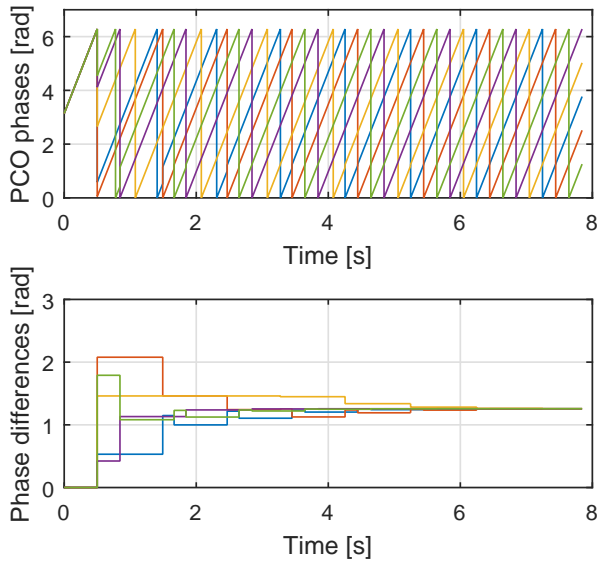}
        \caption{The evolution  of oscillator phases $\phi_k (k=1,\ldots,N)$ (upper panel) and phase differences between neighboring oscillators $\Delta_k (k=1,\ldots,N)$
        (lower panel) under the phase response function $(\ref{eqn:PRC})$. The initial  phases of all nodes were set to the same value.}
        \label{fig:10}
    \end{figure}

\begin{figure}
        \centering
        \includegraphics[width=0.42\textwidth]{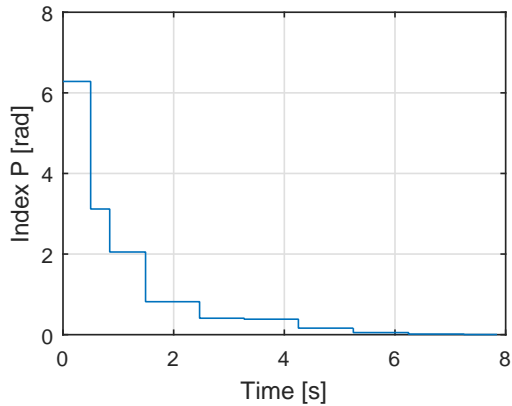}
        \caption{The evolution  of the index $P$ under the phase response function $(\ref{eqn:PRC})$. The initial  phases of all nodes were set to the same value.  }
        \label{fig:12}
    \end{figure}

    We also considered the case where oscillators may have equal initial
    phases.
    We set the initial phases of all oscillators to  $\pi$ and set $l$ in
    $(\ref{eqn:PRC})$ to
      $0.85$. $w$ was set to $2\pi$. Under the mechanism in Sec. IV,
      the evolutions of oscillator phases, phase differences between neighboring oscillators,
       and the index $P$ are given in Fig. $\ref{fig:10}$
       and Fig. $\ref{fig:12}$, respectively. It can be seen that the
       proposed phase
       desynchronization approach can indeed achieve phase desynchronization even
       when
       all PCOs have equal initial phases.

    \section{Conclusions}

    A phase desynchronization approach
    is proposed based on the pulse coupled oscillators.
    Different from existing results which address equal separation of firing time instants and thus
    are subject to uneven spread of phases due to pulse based
    interaction, the proposed approach can achieve constant and even
    phase spread and its performance is guaranteed by systematic and
    rigorously mathematical analysis. Further more, we also proposed
    a mechanism to achieve desynchronization when some oscillators have equal initial phases,
    under which condition almost all existing
    approaches fail to fulfil desynchroniztion.
    Numerical simulations are given to confirm the theoretical results.

\bibliographystyle{unsrt}
\bibliography{abbr_bibli}

\end{document}